# Real-space post-processing correction of thermal drift and piezoelectric actuator nonlinearities in scanning tunneling microscope images

Mitchell P. Yothers, Aaron E. Browder, and Lloyd A. Bumm

*Homer L. Dodge Department of Physics and Astronomy, University of Oklahoma, Norman, Oklahoma 73019, USA*

We have developed a real-space method to correct distortion due to thermal drift and piezoelectric actuator nonlinearities on scanning tunneling microscope images using Matlab. The method uses the known structures typically present in high-resolution atomic and molecularly-resolved images as an internal standard. Each image feature (atom or molecule) is first identified in the image. The locations of each feature's nearest neighbors (NNs) are used to measure the local distortion at that location. The local distortion map across the image is simultaneously fit to our distortion model, which includes thermal drift in addition to piezoelectric actuator hysteresis and creep. The image coordinates of the features and image pixels are corrected using an inverse transform from the distortion model. We call this technique the thermal-drift, hysteresis, and creep transform (DHCT). Performing the correction in real space allows defects, domain boundaries, and step edges to be excluded with a spatial mask. Additional real-space image analyses are now possible with these corrected images. Using graphite(0001) as model system, we show lattice fitting to the corrected image, averaged unit cell images, and symmetry-averaged unit cell images. Statistical analysis of the distribution of the image features around their best-fit lattice sites measures the aggregate noise in the image, which can be expressed as feature confidence ellipsoids.

**I. INTRODUCTION**

Scanning tunneling microscopy (STM) has been an important tool for measuring surface structure for over thirty years. One of the greatest strengths of STM is its real-space imaging of individual atoms or molecules, granting the ability to see surface structures, defects, and features at the sub-nanometer scale.[1, 2] However, its absolute measurement accuracy is not as impressive. For most applications this is not an issue because internal standards are frequently available.

Lattice structures of STM substrates and adsorbate overlayers are usually known very accurately via x-ray diffraction (XRD), low-energy electron diffraction (LEED), etc. While this is helpful for short distances, measurements over long distances can be problematic. On close inspection, STM images often appear slightly warped, especially along the slow-scan direction. Drift due to thermal expansion (thermal drift), as well as piezoelectric actuator nonlinearities (hysteresis and creep), vary too dynamically to be compensated by real-time behavioral models. In addition, position sensors are not sensitive enough to facilitate closed-loop control for the STM at the atomic scale, in contrast to similar techniques like atomic force microscopy (AFM),[3, 4] where nanometer resolution and micron scan ranges are typical. The absolute measurement accuracy of STM images suffers unless these dynamic effects can be compensated. In principle it should be possible to post-correct any STM image that contains regions with known periodic structure, e.g. substrate atomic lattice or adsorbate overlayer. Herein we discuss a post-processing real-space method to use the lattice structure in the image as an internal standard for correcting distortion caused by these dynamic effects. Because it corrects thermal drift, hysteresis, and creep, we call our method the thermal-drift, hysteresis, and creep transform (DHCT). Distortion corrected images from DHCT allow consistent and accurate real-space measurements from STM images, with precision sufficient to register large continuous domains to a lattice.

Some of the earliest work to correct for distortion in STM images involves measuring the average drift effect in an image and applying a voltage ramp to the scan piezoelectric actuators that would compensate for that drift in future images.[1] Since then, some groups have used post-processing methods to correct hysteresis[5, 6] and thermal drift[7-10] individually. These techniques often require additional information, e.g. a piece of the same image scanned with the fast-scan direction orthogonal to the original image.[8] Others have implemented methods to correct distortion from the piezoelectric actuators in real time, e.g. by modifying the image raster scan in the controller frame using behavioral models for the piezoelectric actuators[11-14] and the linearization of long-range hysteresis with software,[15] or by hardware linearization of the piezoelectric actuator response using charge control instead of voltage control.[16, 17] All have proven effective for reducing distortion, although they can introduce new problems.

Post-processing techniques for correcting image distortion have been used successfully for other types of scanning microscopy. Digital image correlation (DIC) has been used to correct scanning electron microscopy (SEM)[18] and atomic force microscopy (AFM)[19, 20] images by determining image distortion caused by drift of the probe with respect to the sample. RevSTEM has been used in scanning transmission electron microscopy (STEM)[21] to correct problems caused by sample drift specific to STEM. Its creators also developed a per-feature post-processing analysis of their previously corrected STEM images.[22] Their methods have been applied to studies of crystallography with STEM.[23, 24] Crystallographic image processing (CIP), although not exactly a distortion correction technique, has been widely applied to electron microscope images[25] and recently expanded to a variety of other microscopies including STM.[26, 27] CIP creates unit cell images by enforcing an assumed symmetry using Fourier filtering and symmetry averaging. This can be advantageous in STM to smooth tip-profile effects which break the natural symmetry of the surface unit cell.

DHCT employs a four-step procedure to correct distortion. 1) Find the location of each image feature (molecule or atom) with sub-pixel accuracy by fitting a 2D Gaussian to the feature's image. 2) Determine the local distortion around each feature by measuring the deviations of that feature's nearest neighbors (NNs) from their expected position. 3) Fit the trend of the local distortion with time to models for thermal drift, hysteresis, and creep. 4) Apply the inverse of the distortion to transform the image and feature data from the controller frame to the sample frame. Distortion correction is thus a recalculation of the in-plane coordinates of each feature and image pixel. It does not alter the z data, thereby maintaining the data integrity. DHCT gives two primary results: a distortion-corrected and calibrated image of the sample surface and the indexed set of features used to determine the distortion. The indexed feature set includes positions, heights, and widths for each feature.

Additional measurements on each feature can also be indexed for later analysis. We also present a set of advanced analysis techniques and their products that are unlocked by accurate distortion correction. Principle among these is lattice fitting to the feature set, which associates a lattice site to each feature. Probability ellipsoids show the statistical deviation of features from their ideal lattice sites. Averaged unit cell images take advantage of the basic translational symmetry of the lattice to generate an image of the unit cell image that is the composite of all the unit cells in the image, generated from the corrected image and the best-fit lattice. In addition to translational averaging, we can create a symmetry-averaged unit cell image (the real-space analog to CIP) that is also averaged over other unit cell symmetries, *e.g.* mirror planes and rotation centers.

Our implementation of DHCT in Matlab and sample STM data for testing are available online at GitHub. The method should work for any sample that fits the criteria discussed herein with some slight modifications to adapt it to different images.

## II. BACKGROUND

Understanding distortion in STM images begins with a discussion of the mechanics and chronology of how STM images are acquired. Fundamentally, image distortion in STM is caused by a discrepancy between the STM controller's model of the probe tip's location (the controller frame of reference) and the probe tip's true location over the surface (the sample frame of reference).[28] We will refer to these as the controller frame and the sample frame.

The position of our STM probe tip is controlled by a piezoelectric tube scanner.[11] The surface region is raster scanned by our STM controller to acquire a square grid of data pixels to be displayed as a digital image. The raster scan in the STM controller frame is shown in Figure 1. The probe tip is scanned at constant velocity along the first row of the scan window, taking measurements that are approximately evenly spaced in time. When the tip reaches the end of the first line, after a brief turnaround time (magenta arrows), it is scanned in the opposite direction along the same row, taking measurements at the same rate. Because the probe tip moves fastest along these rows, this is called the fast-scan direction. We will refer to the initial scan line as the trace (red arrows), and the second scan back along the same line as the retrace (green arrows). After acquiring the entire row (trace and retrace), the tip is moved one pixel width in the slow-scan direction (blue arrows), orthogonal to the fast-scan direction, and is scanned along another fast scan row. This process is repeated until all of the rows in the image are acquired. A pair of images is acquired this way, the trace image and the retrace image.

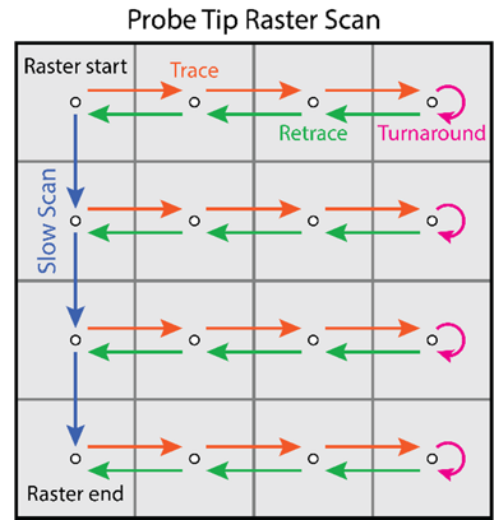

**FIG. 1.** An example STM image raster scan. Each pixel is acquired sequentially with approximately constant time delay between each acquired pixel (open circles). After data acquisition, the samples are displayed as a digital image, which is the grid of pixels arranged corresponding to their position in the raster scan. This example would produce a pair of 4 pixel × 4 pixel images, one from the trace (red arrows) and the second from the retrace (green arrows).

Our STM system uses an open-loop controller, as is typical of most STM systems.[12, 29] The STM controller assumes a linear relationship between the applied voltage of the tube scanner and the resulting motion of the tip. To generate the raster scan for an image data grid, the controller applies a series of voltage steps in both the fast and slow-scan directions, with a significantly higher step rate in the fast-scan direction. Thermal drift of the sample with respect to the probe tip and nonlinearities of the piezoelectric actuator introduce an additional systematic shift of the probe tip position in the sample frame with respect to the controller frame, which is a function of time and voltage. A square data grid in the controller frame becomes a distorted grid in the sample frame.[30] If this data were shown on the regular grid of the controller frame, the image would be distorted. This distortion disappears by transforming the data coordinates from the controller frame into the sample frame. It is helpful to think of distortion as a time-dependent image transformation from the STM sample's frame of reference into the controller's frame of reference. This transformation must be undone to view the STM sample without the effect of distortion.

In our discussion we use time as the independent variable to characterize how distortion varies across the image. This choice is convenient because thermal drift and piezoelectric actuator creep are inherently time dependent, and because time is invariant under the image transformation we use to correct the image. The raster scan used to acquire the image provides a convenient mapping between time and position within the original image. In the controller frame, the time between any two image pixels is independent of absolute position and only depends on the direction and distance between them. In the following sections we will discuss the types of image distortion, the physical origin of the three types of distortions that effect STM images, and how distortion can be measured.

## A. Time-independent Image Transformations

Time-independent affine image transformations can be described by a 3×3 transformation matrix **T** that transforms an image from one coordinate system to another. In this case from the controller-frame image into the sample frame. STM images are topographic maps which give height ($z$) as a function of position ($x$ and $y$). For STM images, it is most natural to express **T** in terms of the fast and slow-scan coordinates of the raster scan $f$ and $s$, respectively, rather than the fixed coordinate frame of $x$ and $y$. Each of the 9 matrix elements can be identified using a pair of indices from the set ($f, s, z$). To understand how the transformation matrix changes the image, we will explore how each transformation matrix element would modify a square image. These transformations are represented visually in Figure 2.

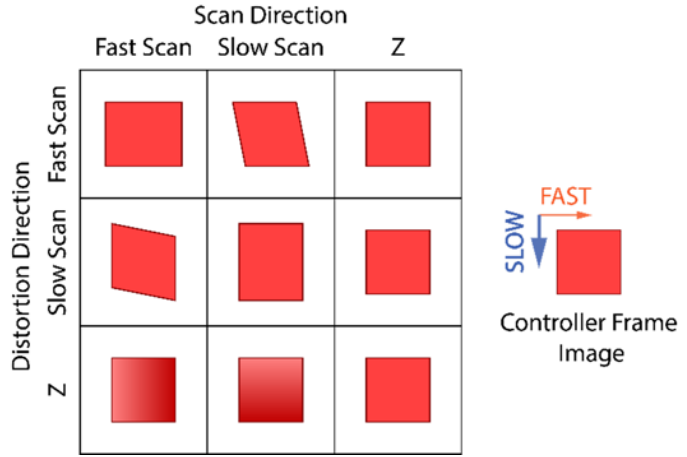

**FIG. 2.** Examples showing how each element of the transformation matrix modifies the scanned region in the sample frame. The controller-frame image is always a square. Displaying the non-square scanned region in a square causes it to appear distorted. The in-plane distortion elements cause stretch and shear, while the z distortion elements cause tilt. Shading (light to dark red) is use to represent changes in z.

The simplest image transformation is the identity transformation, which leaves the image unchanged when applied to the image. In the identity transformation, all of the diagonal elements of the transformation matrix are 1, while the off-diagonal elements are 0.

$$\mathbf{T} = \begin{pmatrix} 1 & 0 & 0 \\ 0 & 1 & 0 \\ 0 & 0 & 1 \end{pmatrix} \tag{1}$$

*1. Stretch*

Stretching an image in a given direction causes distances in that direction to increase by an amount proportional to how much the image was stretched. Compression is a special case of stretching where the stretch amount is negative. The stretch transformation matrix is obtained from the identity transformation matrix by adding the stretch amount in a given direction to the diagonal element of the transformation matrix for that direction.

$$\mathbf{T} = \begin{pmatrix} 1 + S_f & 0 & 0 \\ 0 & 1 + S_s & 0 \\ 0 & 0 & 1 + S_z \end{pmatrix} \qquad (2)$$

$S_f$ and $S_s$ are the stretch amounts in the fast and slow-scan directions, respectively. If $S_f = 0.05$ and $S_s = 0$, the image will be stretched by 5% in the fast-scan direction. $S_z$ can be regarded as a $z$ calibration correction, which is most commonly accomplished using images of an internal standard with known height, such as an atomic step. We will not be considering $S_z$ further, although our method could be extended to correct z calibration in STM images.

*2. Shear*

Shearing an image in a given direction causes points to be offset in that direction by an amount proportional to their distance along a perpendicular direction. Shearing a square image in the plane transforms the square into a parallelogram, which changes the angular relationships between image features. Shear does not change the area of the image. The shear transformation matrix is obtained from the identity transformation matrix by adding the shear amount in a given direction $i$ proportional to distance along perpendicular direction $j$ to the transformation matrix element $\mathbf{T}_{ij}$.

$$\mathbf{T} = \begin{pmatrix} 1 & \mathbf{T}_{fs} & \mathbf{T}_{fz} \\ \mathbf{T}_{sf} & 1 & \mathbf{T}_{sz} \\ 0 & 0 & 1 \end{pmatrix} \tag{3}$$

To shear an image in the fast-scan direction by 2% of the distance in the slow-scan direction, simply set $\mathbf{T}_{fs} = 0.02$ and the other shear matrix elements to zero, then apply that transformation to the image. The $\mathbf{T}_{fz}$ and $\mathbf{T}_{sz}$ matrix elements cause images to shear in proportion the $z$ coordinate, e.g. the z axis is not orthogonal to the $x$-$y$ plane. This introduces a lateral offset between the top of an image feature and the bottom of the feature. Ordinary atomic and molecular corrugation is very small, so the effect is usually insignificant. The most likely manifestation of $\mathbf{T}_{fz}$ and $\mathbf{T}_{sz}$ shear occurs through the larger $z$ motion that results when sample tilt is present. The effect of this is simply an apparent change in the sample tilt, thus is captured and corrected as a part of image tilt. For these reasons we will not be considering the effects of $\mathbf{T}_{fz}$ and $\mathbf{T}_{sz}$ on STM images.

### 3. Tilt

Image tilt is associated with distortion components with the same functional composition as the STM image, $z(x,y)$, which contribute to $\mathbf{T}_{zf}$ and $\mathbf{T}_{zs}$. STM practitioners commonly use plane subtraction to flatten STM images for analysis, a practice generally accepted for STM image processing because image tilt usually carries no useful information.[31] Occasionally higher-order surface subtraction is needed to remove second-order effects of scan curvature and the changing magnitude of creep and thermal drift. One cause of image tilt is sample tilt, the result of the sample surface and the image raster scan being non-coplanar. STM tunneling current feedback forces the raster scan to follow the sample surface, which causes sample tilt to shear the raster scan in the controller frame. Thermal drift and creep components in the $z$ direction also cause image tilt by shearing the raster scan in the controller frame. Piezoelectric tube scanners introduce a curvature to the images because they scan by bending—an effect that increases with increasing offset from the tube scanner's unbent origin.[32] First order effects of raster-scan curvature contribute to $T_{zf}$ and $T_{zs}$.

$$\mathbf{T} = \begin{pmatrix} 1 & 0 & 0 \\ 0 & 1 & 0 \\ T_{zf} & T_{zs} & 1 \end{pmatrix} \quad (4)$$

Plane subtraction of images with sample tilt causes the images to be foreshortened, compressing the image along the tilt gradient. Image tilt due to $z$ components of thermal drift and creep does not lead to image foreshortening and therefore can be fully corrected by plane and higher-order surface subtraction. The error caused by plane subtraction due to foreshortening will be captured in $\mathbf{T}_{ff}$, $\mathbf{T}_{fs}$, and $\mathbf{T}_{ss}$.

## B. Time-dependent Image Transformations and the Mechanisms of Distortion in STM

Thermal drift and piezoelectric actuator nonlinearities cause the elements of the transformation matrix to vary with time. Our correction method solves for $\mathbf{T}_{ff}$, $\mathbf{T}_{fs}$, and $\mathbf{T}_{ss}$ as functions of time using models for thermal drift and piezoelectric actuator nonlinearities. Based on these models, we expect the contribution from $\mathbf{T}_{sf}$ to be small compared to the other 3 elements due to the STM's much higher acquisition rate along the fast-scan direction. Simulated STM images in Fig. 3 separate the effects of thermal drift, hysteresis, and creep and show how they manifest themselves differently depending on their direction with respect to the fast and slow scan directions.

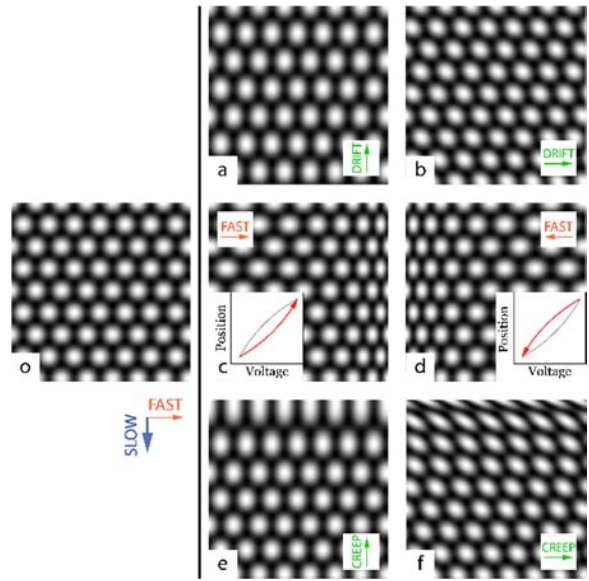

**FIG. 3.** Distortion effects on a simulated STM image of graphite 1 nm × 1 nm, scanned at 50 nm/s, 512 pixel × 512 pixel. (o) The undistorted image. Thermal drift, with a constant velocity of 10 pm/s opposite the slow-scan direction (a) and in the fast-scan direction (b). Hysteresis, the exaggerated effect of exponentially decaying hysteresis in the trace image (c) and the retrace image (d). Piezo creep opposite the slow-scan direction (e) and in the fast-scan direction (f).

## *1. Thermal Drift*

Thermal drift describes the relative motion between the STM probe tip and the sample caused by thermal gradients and time-varying temperature. The mechanical framework of the STM thermally expands and contracts in response to these temperatures. The distance along physical connections between the tip and the sample is 3 to 4 orders of magnitude larger than a typical scan size,[30] so even a small change in temperature will cause noticeable displacement between the tip and sample.

In terms of its effect on the image, thermal drift causes the raster-scan window to move with respect to the sample during the image acquisition. The vector component of thermal drift parallel to the slow scan direction causes the scan lines to be farther apart or closer together in the sample frame, thus stretching or compressing the image features in the slow-scan direction in the controller frame, as shown in Fig 3(a). The vector component of thermal drift parallel to the fast-scan direction causes the start of each subsequent scan line to be systematically and increasingly offset along the fast-scan direction in the sample frame, thus shearing the image features in the controller frame, as shown in Fig 3(b). The vector component of thermal drift parallel to the fast-scan direction also causes the data points in a scan line to be farther apart or closer together in the sample frame, thus stretching or compressing the image features in the fast-scan direction. This latter effect is insignificant in most usable images because its magnitude scales inversely with the fast-scan speed. In contrast, the first two effects scale inversely with the slow scan speed. For a 2048 pixel × 2048 pixel image, this causes the scan velocity along the slow-scan direction to be lower by a factor of about 4096. The magnitude of the $\mathbf{T}_{fs}$ and $\mathbf{T}_{ss}$ matrix elements are therefore the most important for determining the thermal drift.

## *2. Piezoelectric Actuator Hysteresis*

Hysteresis is a property of piezoelectric actuators that causes them to trace out a different (non-linear) voltage-position curve depending on their history.[29, 33] The PZT piezoelectric material used in our STM actuators is a polycrystalline ceramic composed randomly oriented crystallites.[34] Each crystallite is in turn composed of ferroelectric domains each contributing its dipole moment. The allowed orientation of the dipole axes are determined by the material's crystallography. Within that constraint, in the unpoled state the net dipole moment of the domains in each crystallite will be negligible. The random orientation of the crystallites produces a random distribution of allowed polarization axes throughout the material. Poling aligns the domain dipoles within each crystallite so they have a polarization component in the poling direction. The degree to which each domain contributes to the net polarization depends on the angle between each domain's dipole direction and the poling field. When the voltage applied to the poled piezoelectric actuators is stepped, the domains in the piezo become charged. This charge induces strain in the piezoelectric actuator which deforms it, thereby causing motion of the STM probe tip over the sample. The voltage step also brings additional domain dipoles into alignment with the electric field, which decreases the strain and the incremental motion.[35] At sufficiently large voltage, all of the domain dipoles would become aligned and further changes in applied voltage would cause only charging of the domains, resulting in an effectively linear position response to applied voltage. The hysteresis effect depends on the direction of the voltage step rather than the magnitude of the voltage at any given point. Hysteresis results from smaller mechanical displacement per volt when the voltage steps change direction from increasing to decreasing and vice versa, because the change initially drives more domain dipole realignment than charging.

Hysteresis manifests itself as stretching of the image features in the fast-scan direction at the start of each fast-scan line. Due to the STM's periodic motion, the effect of hysteresis in this direction quickly settles into a loop where hysteresis can be observed as a function of time since the scan-line start. An example with exaggerated hysteresis is shown in Fig 3(c) for the trace image (fast-scan direction left to right) and in Fig. 3(d) for the retrace image (fast-scan direction right to left). The effect of hysteresis on each fast-scan line is the same even though each scan-line is acquired at a different time. This is advantageous since all of the features in the image can be used to correct each line, rather than only the features that appear on a single line. The fast-scan hysteresis information is contained in the $\mathbf{T}_{ff}$ matrix element. We do not model hysteresis in the slow-scan direction. Doing so would require information about the probe-tip motion occurring before the image was acquired, thus is not captured in a single image. Although slow-scan hysteresis modeling should be feasible for image sequences, its effect will be combined with thermal drift and creep. Fortunately the lowest-order components of the slow-scan hysteresis will appear identical to drift and creep and be captured in the $\mathbf{T}_{fs}$ and $\mathbf{T}_{ss}$ matrix elements.

### 3. Piezoelectric Actuator Creep

Piezoelectric actuator creep describes the dynamic property of piezoelectric actuator motion to approach the equilibrium position for their bias potential.[29, 33, 36] As explained in the hysteresis section, voltage applied to a piezoelectric actuator causes both strain and domain dipole alignment. Creep is caused by the slow relaxation of the voltage-induced domain dipole alignment to its new equilibrium state over time. As the domain dipoles relax to their equilibrium state, the domains gain additional charge, causing strain that further deforms the piezoelectric.[35] For a step-function voltage change on the piezoelectric actuator, most of the motion occurs as quickly as the mechanical resonance of the system allows. The remaining motion occurs logarithmically with time as the piezoelectric actuator relaxes to its ultimate position.[29, 33]

Large actuator motions, like those that occur during initial sample approach and when selecting a scan window, cause noticeable image distortion due to creep. The effect on the image is similar to thermal drift except with a decaying amplitude. As such, we consider only the $\mathbf{T}_{fs}$ and $\mathbf{T}_{ss}$ matrix elements, for the same reasons as thermal drift. Figure 3(e) shows the stretching caused by the creep vector component parallel to the slow-scan direction. Figure 3(f) shows the shear caused by the creep vector component parallel to the fast-scan direction.

## C. Internal Standard and Region Masks

Typical samples imaged by STM consist of crystalline structures or commensurate overlayers. The deviation of the structure observed in the controller frame from the pristine crystal structures in the sample frame is likely due to the systematic distortion inherent in the controller frame image. In this work, we use the NN distance as the reference value to correct distortion and to calibrate the image. If the exact NN distance is not known, a reasonable guess is sufficient to correct distortion. Strain in the sample also distorts the internal standard. In the case of soft materials, e.g. alkanethiol SAMs, this must be taken in to account near defects, domain boundaries, and step edges. Using region masks to select data away from strained areas yields reliable distortion corrections that can then be applied to the whole image. Region masks facilitate correcting images from patches of the internal-standard structure, e.g. islands in multiphase systems. Performing distortion measurement in real space permits selection of any number of regions with any shape.

**D. Measuring Linear Distortion**

Henriksen and Stipp devised a method to determine linear drift parameters for scanning probe microscopy (SPM) images using the locations of the principle peaks of the Fourier transform.[10] Working with AFM images of graphite, they derived a set of equations that give distortion parameters related to the image transformation matrix elements $\mathbf{T}_{ff}$, $\mathbf{T}_{fs}$, and $\mathbf{T}_{ss}$ from the location of 3 unique peaks of the Fourier transform. Graphite exhibits a close-packed surface structure where each imaged carbon atom has 6 imaged NNs.[37] The Fourier transform of this regular trigonal lattice has six principle peaks. The peak locations depend on the NN distance and the distortion parameters. The authors characterized distortion in terms of the drift velocity components in the slow and fast-scan directions and a homogeneous scaling factor. The drift velocities are related to the matrix elements by the scan velocity in the slow-scan direction. The scaling factor is a constant multiplying $\mathbf{T}$ and refines the instrument calibration. After these parameters are obtained, they are used to apply a linear image transformation. Their method is applicable to determining and correcting time-independent distortion in any kind of raster-scan image, including STM. This method works for distortions that are uniform throughout the image, but hysteresis, creep, and thermal drift vary with time, and therefore not uniform across the image.

**E. Measuring Local Distortion**

By taking local measurements of linear distortion at many points throughout the image, we can determine the trend of the distortion with time. Initially we set out to do this by adapting the Henriksen-Stipp method using a sliding FFT. In this approach, a small window is moved across the image, using the FFT of each window to determine the local distortion at the center of the window. However, the sliding FFT approach has two drawbacks. First, distortion information is averaged over the FFT window. Reducing the window size to increase the spatial resolution comes at the expense of the k-space resolution, making it difficult to determine peak locations. Second, images with "interesting" features (like defects and boundaries) add additional Fourier components that make measurement of local distortion more difficult when the window contains those features.

As a solution to the problems with the sliding FFT method, we use the locations of image features in real space as the basis for DHCT. For graphite, since the imaged surface is a trigonal lattice of carbon atoms, the NNs should lie on a circle and be equidistant from each other and from the central atom, just like the peaks of the Fourier transform. By applying our method to each feature's NNs, we determine the local drift parameters at each feature location in the image. Performing the analysis in real-space, we avoid the problems of the FFT and we add the ability to use region masks to select regions of the surface for analysis. The drift velocities are now determined as functions of time to account for creep and time-varying thermal drift, and the scale factor has been adapted to measure fast-scan hysteresis as a function of time since the beginning of the scan line. It should be noted that hysteresis depends on voltage rather than time. Voltage is mapped to time for convenience, because the voltage is stepped at a constant rate in the fast scan direction.

### III. MATERIALS AND METHODS

The instrument used for these measurements is a beetle-style STM. The STM is controlled by an SPM100 from RHK Technology with a tunneling current measured using an Axon CV4 current amplifier. XPM Pro software, also from RHK Technology, is used to monitor and modify the scan during acquisition and save the scan results. STM imaging was performed in dry $N_2$ at room temperature. Tips were prepared from Pt-Ir (80:20) wire by clipping with wire cutters. Computation was performed using Matlab 2016a on a Dell 7910 with dual 4-core Dual Intel Xeon processors (E5-2637), 128 GB RAM, and an NVIDIA Telsa K40 GPU. Sections of the code are optimized for parallel processing or the GPU, as was expedient.

We have selected two model systems to test the technique: highly-ordered pyrolytic graphite (HOPG) and self-assembled monolayers (SAMs) of 1-decanethiol on Au(111). Our HOPG was a ZYB-grade sample obtained from NT-MDT. Before imaging, the graphite sample was freshly cleaved to expose a pristine surface. The graphite was imaged at −60 mV sample bias and 160 pA tunneling current in constant current mode. Au(111) substrates deposited on mica obtained from Agilent Technologies were reused from previous experiments by annealing with a hydrogen torch. After flame annealing, the sample was put into an airtight container with 5 μL of 1-decanethiol and purged with nitrogen for 5 minutes before being sealed. The sealed container was heated in an oven at 100 °C for 4 hours, then removed and allowed to cool to room temperature. The sample was then rinsed with toluene, then ethanol, and blown dry with $N_2$. The SAMs were imaged with the STM at −1.00 V sample bias and 1.0 pA tunneling current in constant current mode. 1-decanethiol 99% was obtained from Sigma-Aldrich and used without further purification.

## IV. TECHNIQUE

### A. Pre-processing

Our analysis begins by reading the .sm3 or .sm4 STM image proprietary file format of RHK Technology into Matlab as a pair of matrices of $z$ coordinates (the trace and retrace image) and the scan parameters from the file header. Once imported, analysis should be independent of the data acquisition software. Plane subtraction fitting is performed by selecting multiple continuous crystal domains with region masks, then the slope of the best-fit plane is found for all of the regions simultaneously and a plane with that slope is subtracted from the image. The tilt can also be manually adjusted to account for each individual sample, but corrections from the best-fit plane are usually small. Reduction of scan-line noise due to sudden tip changes is performed using the linear-regression fitting method from Fogarty et. al.[31] This method works by solving for and subtracting frequency noise in the slow-scan direction that is not correlated with image features.

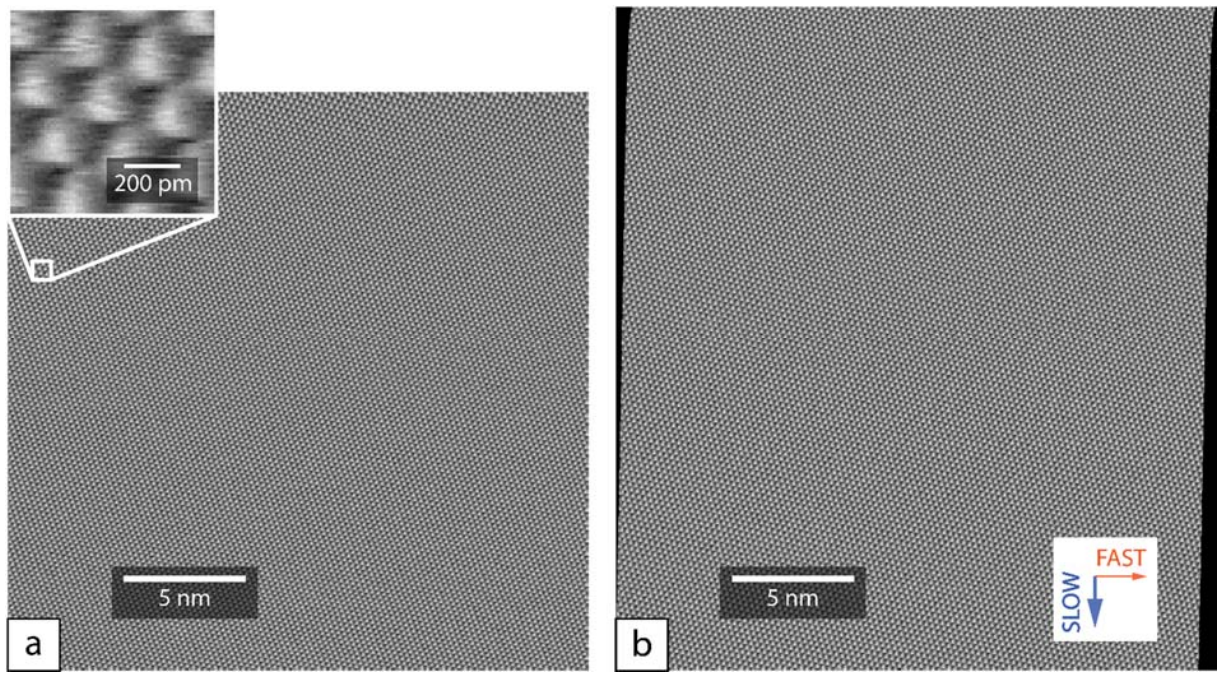

**FIG. 4.** (a) A 2048 pixel × 2048 pixel STM image of graphite after pre-processing, acquired at −60 mV sample bias and 160 pA tunneling current. The image is nominally 25 nm × 25 nm scanned at 100 nm/s. The inset is 800 pm × 800 pm section, outlined in white, is magnified in the inset to show the surface texture. (b) The same image with the total drift correction applied to it. Note that the image is no longer square, the region shown more accurately shows the shape of the region scanned on the surface.

Only images with high quality can be used. After this pre-processing, the $z$ coordinates of the topographic image are no longer modified. Subsequent distortion correction modifies the fast and slow-scan coordinates of each pixel according to the determined distortion, which can be applied to the processed data or the raw data, whichever is more suitable for the application.

We chose to use graphite as the primary sample for this work, both because it is widely studied and due to the rigid nature of the surface structure, compared to soft systems such as alkanethiol SAM surfaces. An image processed as described above is shown in Fig. 4(a) with its distortion-corrected image Fig. 4(b). The image selected was free of large probe tip changes and the lattice was imaged coherently. As you can observe from the inset in Fig 4(a), the probe tip is not ideal, but nevertheless produced stable atomic-resolution images. The secondary example is an STM image of a decanethiol SAM on Au(111), which includes a substrate step edge, two terraces, and two large SAM structural domains, (Fig. S1). Region masks are needed to limit the distortion measurement to regions of SAM domains away from the domain boundaries and step edge.

**B. Initialization**

Our implementation of DHCT begins with the image to be drift corrected as a 2D matrix of $z$ coordinates, and a few parameters describing the image. The distance between any two adjacent pixels (in meters), the time between two sequential pixel acquisitions (in seconds), a pair of strings indicating the slow and fast-scan directions, and the known NN feature distance (in meters). Except for the known NN distance (internal standard), the other values should be the image parameters recorded with the image. These parameters let us determine the scan controller frame of reference as accurately as possible. If a trace and retrace image are to be analyzed simultaneously, both images and the fast-scan direction for each image must be supplied, the other parameters are the same for both images. Optionally, a logical region mask the same size as each image can be included, which will make our software use only positively-masked features for the distortion analysis. An image consisting of multiple structural domains and atomic terraces, are shown in Fig. S1(a), as well as the region masks selecting them (Fig. S1(c)). Masks should be chosen to exclude surface defects, domain boundaries, step edges, or regions with different or unknown structure. The unmasked areas will still be corrected for distortion, but those regions will not be used to calculate the distortion used in the correction. A good correction requires representative regions across the entire image, and the best corrections use as much of the image as possible.

The equations in the following sections assume that $x$ is the fast-scan direction and $y$ is the slow-scan direction for the sake of using simplified coordinate nomenclature. The coordinates that correspond to the fast-scan and slow-scan directions are defined in the initial image parameters passed to the DHCT software.

## C. Feature Indexing

We begin with a standard pixel-based cross-correlation. A kernel image of a feature that has the same approximate size and shape as the image features is generated. We have chosen a radially symmetric 2D Gaussian as a kernel for the two model systems discussed here, but the kernel could be any size or shape. Our kernel is a $(4\sigma + 1) \times (4\sigma + 1)$ square of pixels, where $\sigma$ is chosen to the nearest half-integer based on the size of the image features in pixels. The image is cross-correlated with the kernel and a threshold is applied to select regions with a correlation > 0.1 (regions that look like the Gaussian). An example cross-correlation image for the inset from Fig. 4 is shown in Fig. 5. Each individual region, selected by the cross-correlation and isolated using a watershed function, is fit in the least-squares sense to a more accurate shape model. For this fitting we used this more general 2D Gaussian:

$$g(x,y) = A \exp\left(\frac{-[(x-x_0)\cos\theta + (y-y_0)\sin\theta]^2}{\sigma_1^2} + \frac{-[(x-x_0)\sin\theta + (y-y_0)\cos\theta]^2}{\sigma_2^2}\right) + z_0, \quad (5)$$

where $x_0$, $y_0$, and, $z_0$ are coordinates of the center of the base of the Gaussian, $A$ is its amplitude, $\sigma_1$ and $\sigma_2$ are its standard deviations in 2 orthogonal directions, and $\theta$ is the rotation angle. The height of the center of the feature is $A + z_0$. This 2D cross-correlation and Gaussian fitting was inspired by a similar technique in STEM by Sang, Oni, and LeBeau.[22] We save all of these parameters for each feature in the image with an index to keep track of features for subsequent analysis.

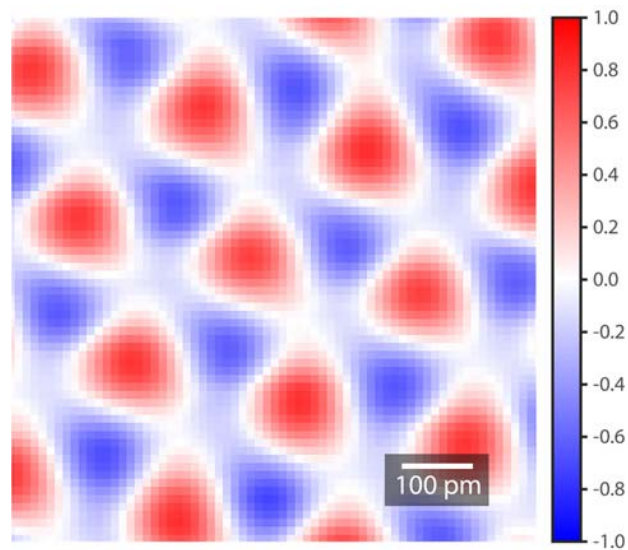

**FIG. 5.** The cross-correlation of the inset part of Fig. 4 with the chosen Gaussian kernel. Red (blue) indicates positive (negative) correlation. A threshold at correlation +0.1 combined with a watershed function on the cross-correlation data selects pixel regions of the surface to analyze with Gaussian fitting.

### D. Time Assignment

Time isn't recorded explicitly for the original image in our STM, but assigning a time to the image features is necessary for determining the distortion trends with time. Because the sampling rate of the STM is approximately constant, we can assign a time to each image pixel based on their known chronological order from the raster scan and from the pixel acquisition rate determined by the scan velocity. In principle, the actual acquisition time of each image pixel in the original STM image could be recorded in a separated clock channel, but the approximate pixel time calculated from the scan speed is sufficient. Image features are not localized to individual pixels, they are imaged by many pixels over multiple raster-scan lines. For example, in Fig. 4 each feature is imaged by approximately 111 pixels over 12 raster-scan lines. The location of each image feature is derived from the 2D-Gaussian fit to pixels across many raster-scan lines. For our analysis we assign a time to each feature based on the time of the image pixel nearest to its best-fit location.

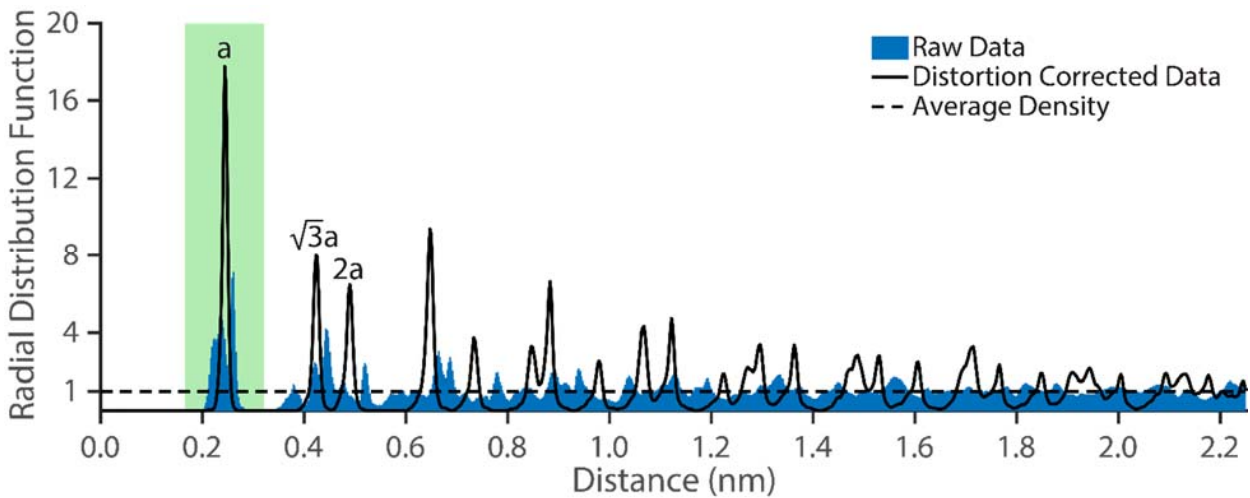

**FIG. 6.** The radial distribution function of image features is used to choose NNs. The NN distance for the imaged carbon atoms in graphite, $a = 2.46$ Å, is used to classify the first 3 peaks of the density correlation (filled blue region). Gaussians centered at $a$, $\sqrt{3}a$, and $2a$ are fit to the unprocessed data to determine the NN distance. All of the features in the green shaded region are used as NNs in the analysis. After the correction, the peaks form the expected radial distribution function for a trigonal lattice (solid black line).

### E. Finding Nearest Neighbors

The spatial relationship of the features is determined by first calculating the distance between every pair of features. The radial distribution function of the indexed features (Fig. 6) shows the aggregate probability density of finding a feature a certain distance away from any other feature, normalized to unity at infinite separation. The first and highest maximum of the radial distribution function occurs at the NN distance, which we will call $a$. To determine $a$ as accurately as possible for the supplied image, Gaussians centered at $a$, $\sqrt{3}a$, and $2a$ (the nearest, next-nearest and next-next-nearest neighbor distances) are fit to the radial distribution function. The value of $a$ is determined accurately enough this way to pick each feature's NNs. The NNs of each feature fall within the shaded box in Fig. 6, within ±35% of $a$. This technique works well for finding NNs, but due to peak broadening at larger feature separations and overlapping peaks, more distant features cannot be reliably selected with the radial distribution function of the raw data. After identifying all of the NNs, we select only the features with the correct number of NNs, e.g. graphite's six imaged NNs.[37] This selection excludes features near defects, mask region edges, the edges of the image, and feature finding errors. Usually there are zero errors, but occasionally an image feature will be found twice (due to scan noise), causing both of those features and all of their NNs to have too many NNs. In a future extension of this work, we plan to investigate features with less than the full number of NNs, which will allow inclusion of features at and close to boundary edges.

## F. Determining Local Distortion

We use each molecule's NNs to determine the local distortion using points chosen in real space. For each of the sets of NN locations, we determine the best-fit ellipse from the minimization of this sum:

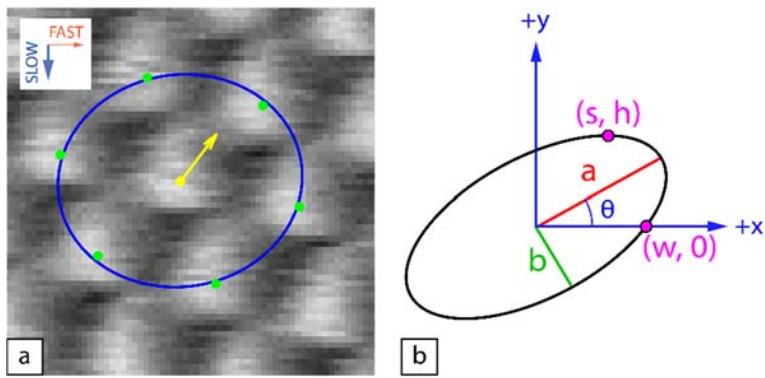

**FIG. 7.** (a) The inset from Figure 4(a), with the central feature's NNs (green), the best-fit ellipse to those NNs (blue), and the drift velocity vector that would cause a circle to be distorted to that ellipse (yellow) overlaid. We determine the drift velocity at each feature location on the surface, thus the velocity determination happens many times. (b) A diagram of an ellipse showing its semi-major axis $a$ and semi-minor axis $b$, with its semi-major axis at an angle $\theta$ from the +x axis, and the derived parameters $s$, $h$, and $w$.

$$\sum_{i=1}^{N}\left[\left(\frac{(x_i-x_0)\cos\theta-(y_i-y_0)\sin\theta}{a}\right)^2+\left(\frac{(x_i-x_0)\sin\theta+(y_i-y_0)\cos\theta}{b}\right)^2-1\right]^2, \quad (6)$$

where $(x_i, y_i)$ are the coordinates of the $i$th of $N$ NNs (Fig. 7(a)). The parameters of the best-fit ellipse are the center of the ellipse $(x_0, y_0)$, the semi-major axis $a$, the semi-minor axis $b$, and the angle to the semi-major axis $\theta$. We turn these measurements into three useful derived parameters: the shear offset and height of the highest point on the ellipse from the center $(s, h)$, and the width from the center of the ellipse to the $x$ intercept $w$, Fig. 7(b). The height $h$, determined from the ellipse equation by setting $\frac{dy}{dx} = 0$, solving for $y$ and choosing the positive solution, is given by

$$h = \sqrt{(a \sin\theta)^2 + (b \cos\theta)^2} \; . \tag{7}$$

The shear offset $s$ is related to $h$ by solving the original ellipse equation for $x$ and evaluating it at $y = h$,

$$s = \frac{-h \sin\theta \cos\theta (a^2 - b^2)}{(a \sin\theta)^2 + (b \cos\theta)^2} \; . \tag{8}$$

The width $w$ is one of the axes of the ellipse before undergoing the shear transformation. Since the shear transformation preserves the area of the ellipse, the area of the best-fit ellipse $\pi ab$ and the area of this un-sheared ellipse $\pi wh$ must be the same,

$$w = \frac{ab}{h} \; . \tag{9}$$

The scale correction, a ratio of input to output intermolecular distance, is a unitless constant derived from $w$ and $r_0$,

$$S = \frac{w}{r_0}, \tag{10}$$

where $r_0$ is the user-supplied value of the NN distance. The shear offset is converted into a drift velocity by multiplying by the scan velocity in the slow-scan direction $v_0$

$$D_x = v_0 \frac{s}{r_0} \; . \tag{11}$$

The height is similarly converted into a drift velocity, but we subtract $v_0$ to remove the effect of the real net scan velocity in that direction

$$D_y = v_0 \frac{h}{r_0} - v_0 = v_0 \left(\frac{h}{r_0} - 1\right). \tag{12}$$

We assign the distortion parameters $D_x$, $D_y$, and $S$ determined from each feature's NNs to that feature for determining the distortion trend with time from the distortion models. An example showing the location of features, their best-fit ellipse, and the resultant drift vector are shown in Fig. 7(a). Our model of the NN structure does not assume the NNs should be uniformly spaced around the central feature—only that they are at a uniform distance (fall on a circle). This model should work for disordered systems as well, provided that identifiable features can be resolved and have a well-defined mean NN distance.

### G. Determining Distortion Trends

To visualize the trend for hysteresis in the fast-scan direction, we plot $S$ as a function of time since the beginning of the scan line (Fig. 8). We find the best-fit curve to the data that obeys the quartic polynomial function:

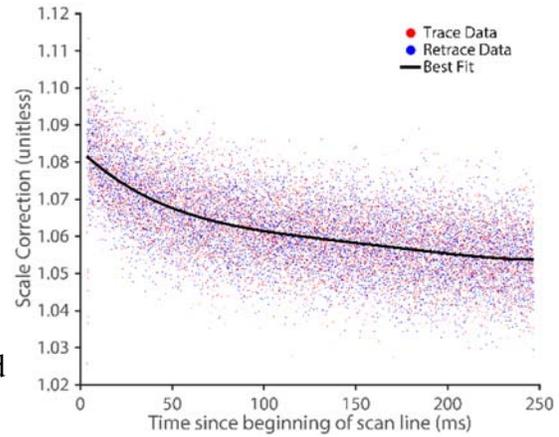

**FIG. 8.** Curve of best fit following our hysteresis model to a plot of scale correction vs time since the beginning of the fast-scan line.

$$S(t) = \sum_{n=0}^{4} a_n t^n, \tag{13}$$

where $a_n$ are the determined coefficients of the $n$th order term of the polynomial. A polynomial function was chosen for its simplicity, since fully modeling the hysteresis curve requires many free parameters. We have also explored using the exponential as a fitting function, but find it does not capture the steepness of the initial part of the data as well. Liu, et. al. were able to fully model a hysteresis loop using a fractional-order Maxwell resistive capacitor model with 21 free parameters,[35] where our polynomial uses only 5. A best-fit curve to the hysteresis data for one of our images is shown in Fig. 8.

To visualize the trends of drift and creep, we plot the drift velocity in both the slow and fast-scan directions as a function of time (Fig. 9). We find the best-fit curve to the drift velocity[33] that obeys the function

$$D_{x,y}(t) = v_{x,y} + a_{x,y}t + \frac{C_{x,y}}{(t+t_0)}, \tag{14}$$

where $v_{x,y}$ is the drift velocity from thermal drift, $a_{x,y}$ is the drift acceleration from thermal drift, and $C_{x,y}$ is the creep function amplitude. The effective time since the impulse that caused the creep, $t_0$, is the only fitting parameter shared by both the $D_x$ and $D_y$ fits. The $x, y$ subscripts indicate orthogonal components of the drift vector in the $x$ and $y$ direction. Including the drift acceleration term significantly improves the image correction. We note that most of the published thermal drift corrections assume a constant velocity thermal drift, which results in a simple linear image correction (affine transformation).[1, 7-10] This can be attributed to the added complexity of measuring and applying a nonlinear correction, which we are already obliged to do because of the piezoelectric actuator nonlinearities. We are using the log(*t*) style creep model where the position offset is related to log(*t*/*t₀*), so the creep velocity depends on its derivative, or $(t + t_0)^{-1}$. Figure 9 shows a plot of measured local drift velocities in both scan directions with their best-fit drift velocity curve according to this model.

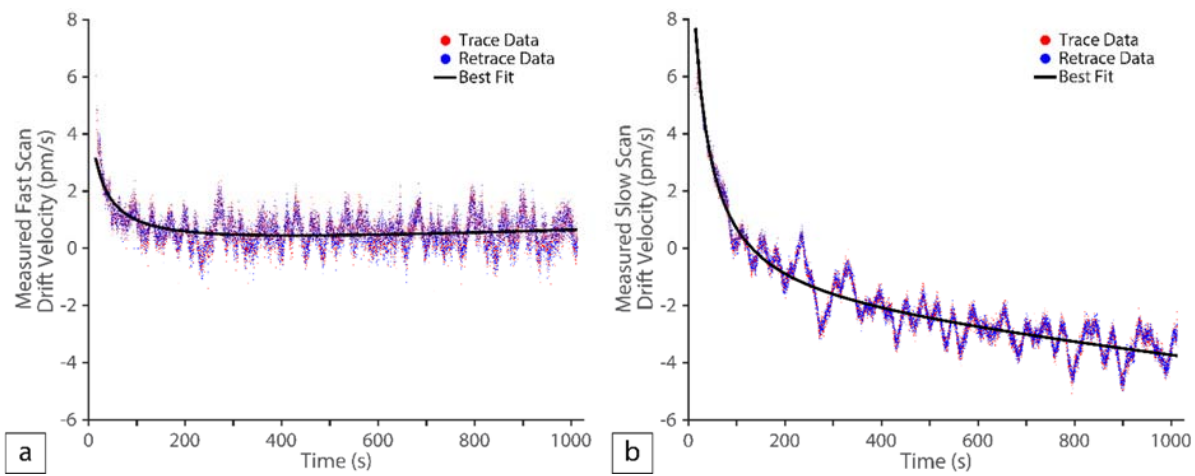

**FIG. 9.** Curves of best fit following our thermal drift and creep model to a plot of measured drift velocity vs time since the beginning of the raster scan. Measured drift velocity is proportional to image distortion matrix elements. Fluctuations around the best-fit curves are likely due to changing tip conditions over the scan time.

In this procedure we simultaneously fit distortion around each feature in the feature sets from the trace and retrace images to the 11 parameters of our distortion model. Considering that this image pair is composed of $2^{23}$ pixels containing a combined total of 22,744 features, 11 parameters should not over fit the data. The deviations about the best-fit curve in Fig. 8 are uniform across the line because the time scale is short. In contrast the scatter about the best-fit curves in Fig 9(a) and (b) vary on a time scale of minutes and are correlated in time. We attribute the residual scatter to changes in the probe-tip.

## H. Applying Image Correction

Combining all of the systematic distortions of the STM image, the *x-y* coordinates of the features and image pixels can be mapped from the controller frame into the sample frame. We integrate the best-fit curves for the distortion model (eqs. 13 and 14, Figs. 8 and 9) to determine displacement from effective drift velocity and evaluate the integral at each time point to find the controller-frame-to-sample-frame displacement at that point. By adding this to its location we obtain the corrected coordinates. The original image is composed of square pixels on a regularly spaced grid, while the corrected image is just a set of (*x, y, z*) data. It is important to understand that this distortion correction does not alter the measured topography (*z* coordinate), so the STM data integrity is preserved. However, display of the corrected image as a standard raster graphic does require resampling the corrected image onto a regularly spaced array. This can be accomplished using the Matlab griddata() function. Black bars on the side of these new images fill the image rectangle where no data is available rather than cropping data out of the image. The results of DHCT can be seen in Fig. 4(b) for graphite and in Fig. S1(b) for a SAM on Au(111).

## V. RESULTS AND DISCUSSION

We have measured and corrected the distortion due to thermal drift, hysteresis, and creep of the STM image using NN image features. We have also assigned feature indices to each surface feature that we use to track information about them, e.g. their relative position in $x$, $y$, and $z$; information about their shape; and which features are NNs. This NN-based correction is remarkable in its ability to restore long-range order well enough to determine a best-fit lattice to the data. Initially we believed the correction would be further improved by performing a second pass of DHCT using more distant features in place of the NNs because it should increase the signal to noise ratio of the distortion measurement. When we tested the idea using graphite images, we did not observe any noticeable improvement. We attribute this to the strength of using a large number of correlated measurements for the distortion measurement, despite using NNs.

### A. Lattice Fitting

Determining the best-fit lattice to the surface structure is desirable for analysis of the image features. We use continuous regions of the masks defined earlier that cover one crystal domain and fit the feature locations selected by that mask to 5 parameters: the origin of the lattice $(x_0, y_0)$; the angle of the first lattice vector from the $x$ axis $\theta$; and a scaling factor in the $x$ and $y$ directions that allows the lattice to fit. If two images were supplied, e.g. a trace-retrace pair, a small offset $(\Delta x, \Delta y)$ is needed between images to allow the identical features to line up. To avoid becoming trapped in a local minimum, our lattice fitting procedure requires the fit region of the image to agree with the initial-guess lattice to better than one half of the lattice constant over the whole region before fitting. In the graphite image in Fig. 4, the accuracy needs to be better than 0.5 × 246 pm over the 25 nm image (better than 0.5%).

To perform the whole-image lattice fit, the lattice angle is guessed by finding the peaks of the angular distribution function of NNs (Fig. S2). Next, a lattice is fit to the features using least squares optimization of the lattice angle, *x-y* offset, and lattice vector magnitude. Finally, each feature is assigned lattice parameters from the fit: two integers that indicate how many lattice vectors the feature is away from the origin. Now registered to a lattice, we can determine how each feature's location is related to each other feature's location, how well each feature's location agrees with the whole-image best-fit lattice, and a variety of other useful measurements. Figure S3 is a heat map of the magnitude of the deviation of each feature's location from its whole-image best-fit lattice site. The striations along the fast-scan direction show that the correlation of the features to a lattice is better over short time and worse over long times, which is consistent with the distortion data shown in Figs. 8 and 9.

## B. Averaged Unit Cell Images

The profiles of atoms and molecules in samples imaged by STM can be obscured by probe-tip fluctuations and by electronic noise. This is particularly true with room-temperature STM imaging, where smaller raster-scan sizes with high-pixel density are not practical, in contrast to imaging with low-temperature ultra-stable STMs. Images that are simultaneously small, slow, and high-resolution are severely impacted by thermal drift and creep in a way that is likely too dramatic to correct. Instead, imaging a large number of unit cells at low resolution affords the opportunity to combine the measurements from each unit cell into to an averaged unit cell image. The room temperature graphite image in Fig. 4 contains 11,372 unit cells, with ~350 pixels per unit cell in the original image. The large number of low-pixel density unit cell images can be averaged together to create a high-resolution translationally averaged unit cell image. This is possible after accurate distortion correction using DHCT and subsequent lattice fitting to identify the boundaries of each unit cell in the image. Subsets of the unit cells can also be selected. For example the location, height, and shape characteristics of each image feature can be used to select the unit cells that are averaged together, e.g. the average of unit cells with

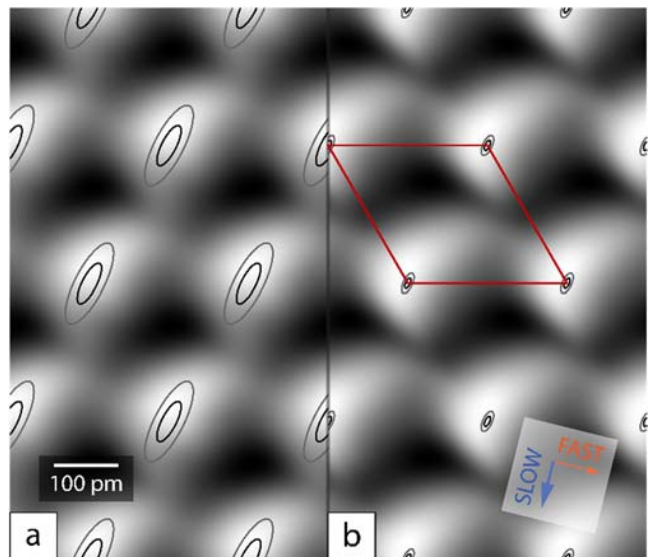

**FIG. 10.** Translationally averaged unit cell image for the carbon atoms on the graphite surface (grayscale images) with the surface unit cell (outlined in red) and 1σ and 2σ confidence intervals of atom locations (outlined in black) overlaid. (a) Data extracted using the whole-image best-fit lattice. (b) Data extracted using the local unit cell fit. The increased spread of the atoms in the slow-scan direction is attributed to the much higher sensitivity to noise, because the sampling rate of neighboring pixels in the slow-scan direction is a thousand times less frequent than in the fast-scan direction.

presence or absence of a certain defect. Averaged unit cell images have proven useful for analysis of STM images of a variety of materials.[9, 38, 39]

The averaged unit cell images of graphite extracted from Fig. 4(b) are displayed in Fig. 10 as the grayscale images. The average unit cell images constructed in two different ways. Figure 10(a) uses the whole-image best-fit lattice to the entire feature set to define all the unit cell boundaries, i.e. a single continuous lattice. Figure 10(b) is a local unit cell fit, using the best-fit unit cell to the four corner features, thus taking advantage of the better short-range correlation. The unit cell orientation is constrained to the angle determined by the whole-image best-fit lattice, because noise will not rotate the image. Figure S4 shows the best-fit unit cells on the image for both methods. Eight averaged unit cell images are tiled in each part of Fig. 10 for ease of visualization and to demonstrate that the boundaries stitch together without discontinuities. One of the unit cells is outlined. After choosing a set of surface unit cells to average, the unit cells are sampled into an appropriately sized grid via a linear transformation of the part of the image that contains the unit cell. The cells are then simply averaged together. The images exhibit a lower symmetry than expected for graphite due to convolution with a probe-tip profile that has a different symmetry. This is especially evident from the profile of the carbon atoms and the three-fold hollow sites. The image derived using the local unit cell fit reveals the character of the probe-tip profile more strongly compared to that derived from the whole-image best-fit lattice.

### C. Feature Confidence Ellipsoids

The image features do not fall perfectly on their best-fit lattice site but are instead distributed around that site. This is the aggregate effect of residual distortion left after DHCT due to higher-order thermal-drift, hysteresis, and creep effects or lattice fitting mismatch; noise in the STM due to the current amplifier, control electronics, and analog to digital converter; noise due to probe-tip fluctuations; and environmental noise from vibration. The $1\sigma$ and $2\sigma$ in-plane confidence ellipses are overlaid on the averaged unit cell image in Fig. 10. The measured $1\sigma$ confidence ellipsoid is stretched in the slow-scan direction. The ellipsoids are largest for the whole-image best-fit lattice ($\sigma_{xy}$ = 11.2 pm × 34.6 pm; $\sigma_z$ = 3.9 pm) and approximately four times smaller using the local unit cell fit ($\sigma_{xy}$ = 3.5 pm × 8.4 pm; $\sigma_z$ = 3.9 pm). We attribute the difference to the longer time scale of feature correlations in the slow-scan direction and the greater influence of probe tip stability. It is instructive to compare this uncertainty to the room temperature rms vibration amplitude of graphite, which is 6.3 pm in plane and 10 pm out of plane.[40, 41] The lattice vibrations occur at $10^{14}$ Hz, much too fast to be imaged directly by STM, nevertheless the comparison provides some perspective on the relative magnitudes of the positional uncertainly measured by STM and the positional uncertainty due to thermal vibration.

## D. Symmetry-Averaged Unit Cell Images

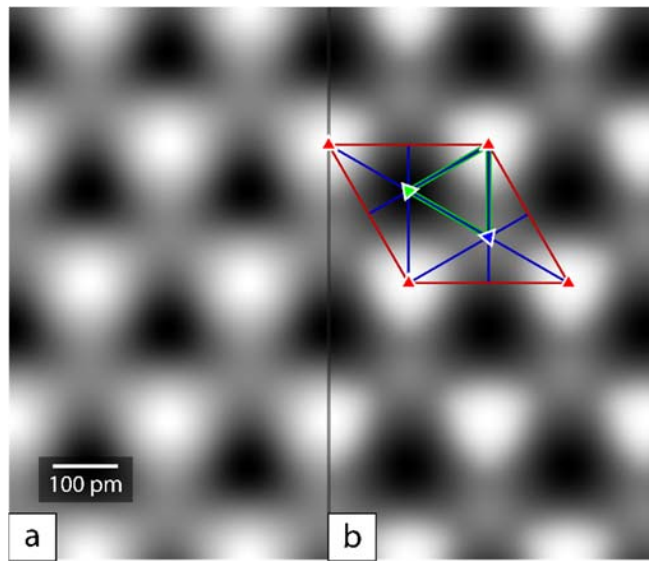

In addition to the translational symmetry that defines the surface unit cell, many structures have additional symmetry. Graphite(0001)'s surface unit cell has p3m1 symmetry. The additional step of averaging over the unit cell's symmetry (e.g. rotations and reflections), enforces the symmetry of the surface unit cell (Fig. 11). Symmetry averaging also smooths over image artifacts with different symmetry, e.g. the probe-tip profile.

**FIG. 11.** Symmetry-averaged unit cell image for the carbon atoms on the graphite surface (grayscale image) with the surface unit cell (outlined in red) and asymmetric unit (outlined in green) overlaid. In addition to translational averaging, this unit cell was also averaged over rotations by 120° about the 3-fold rotation axes (triangles) and reflection about the mirror plane (in blue). Averaging the image this way ensures that the unit cell has the same symmetry as the lattice that generated it. (a) Data extracted using the whole-image best-fit lattice. (b) Data extracted using the local unit cell fit.

## VI. CONCLUSION

We have implemented the thermal-drift, hysteresis, and creep transform (DHCT) as a method for removing distortion from STM images which contain regions of known structure that can serve as an internal standard. The method models the local distortion caused by thermal drift and the piezoelectric actuator nonlinearities, hysteresis and creep. We have demonstrated DHCT on both graphite (HOPG) and decanethiol SAMs on Au(111). The image correction only modifies the ($x$, $y$) coordinates of the acquired image and leaves the $z$ data unchanged. The corrected coordinates also can be applied to the raw $z$ data (before preprocessing), leaving all information in the original image available for detailed analysis. DHCT could be extended beyond trigonal lattices with suitable modifications to the local distortion measurement. The other routines from DHCT should be independent of the surface lattice structure. As a byproduct of the correction, the image scale is calibrated and essentially every feature (molecule or atom) in the image has been located and individually indexed. Each feature's properties are stored together, allowing additional analysis. Averaged unit cell images and confidence ellipsoids can be then extracted which provide high-resolution images of the surface unit cell with a statistical measure of the distribution of the features from their best-fit lattice.

**SUPPLEMENTARY MATERIAL**

See supplementary material for DHCT corrected decanethiol SAM STM image, angular distribution function for graphite NNs in the corrected image, heat map of feature deviations from whole-image best-fit lattice sites, graphite images showing overlay of the whole-image best-fit lattice and local unit cell fit.

**ACKNOWLEDGEMENTS**

Financial support from Homer L. Dodge Physics, and the OU College of Arts and Sciences is gratefully acknowledged. The author's also thank the Carl. T. Bush Fellowship (MPY) and the NSF REU PHY-1062774 (AEB) for support. We gratefully acknowledge the support of NVIDIA Corporation with the donation of the Tesla K40 GPU used for this research.

## Supporting Data: Real-space post-processing correction of thermal drift and piezoelectric actuator nonlinearities in scanning tunneling microscope images

Mitchell P. Yothers, Aaron E. Browder, and Lloyd A. Bumm
*Homer L. Dodge Department of Physics and Astronomy, University of Oklahoma, Norman, Oklahoma 73019, USA*

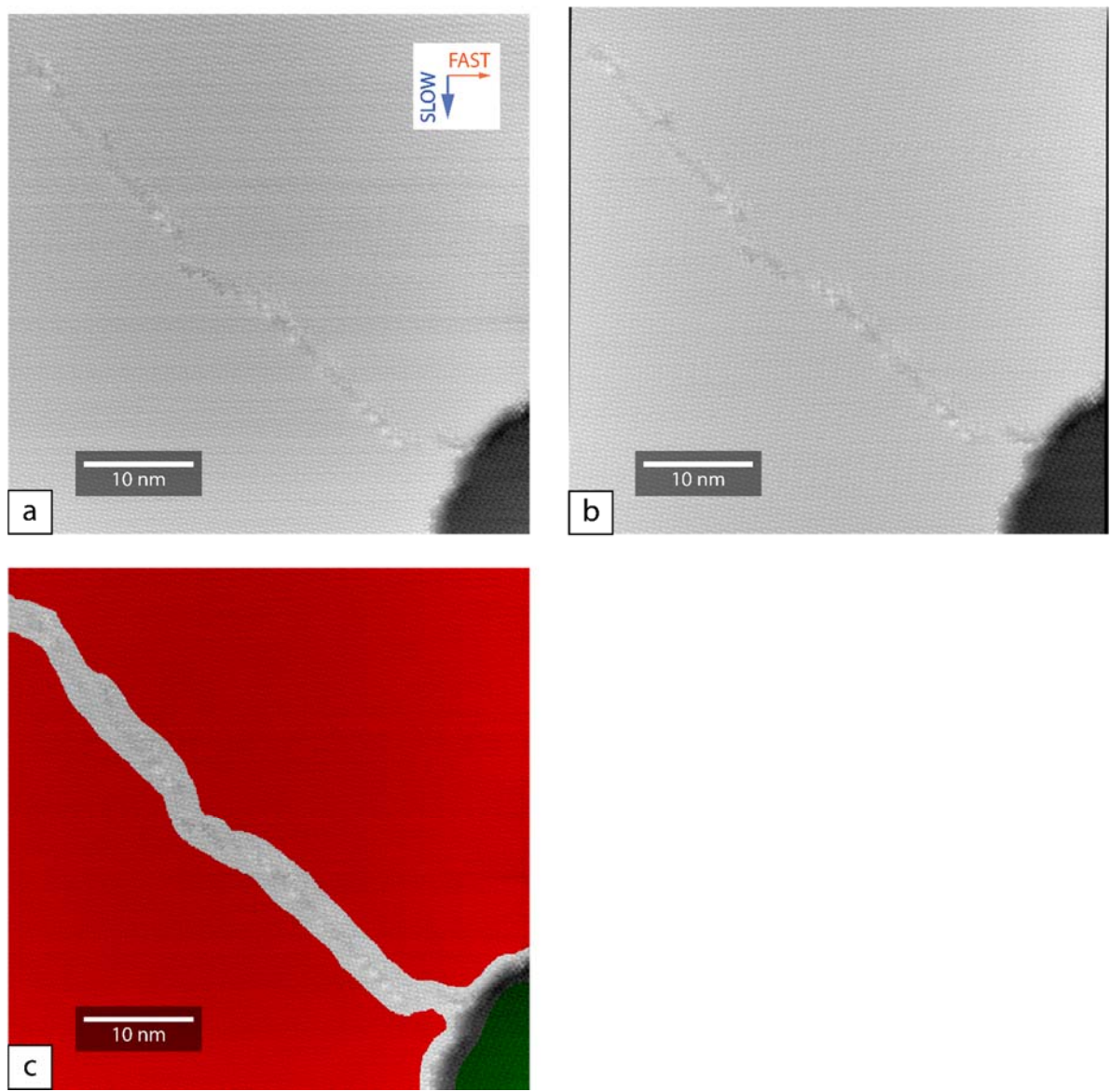

**FIG. S1.** (a) A 1024 pixel × 1024 pixel STM image of a 50 nm × 50 nm region of a decanethiol SAM on Au(111) acquired at −1.0 V sample bias and 1.0 pA tunneling current. The region has structural domain boundary breaking the image into distinct sections. (b) The same image after the distortion correction is applied to the entire image. (c) The same image with three region masks defining the areas to include in the distortion analysis. The upper gold terrace masks are highlighted in red, while the lower terrace is in green.

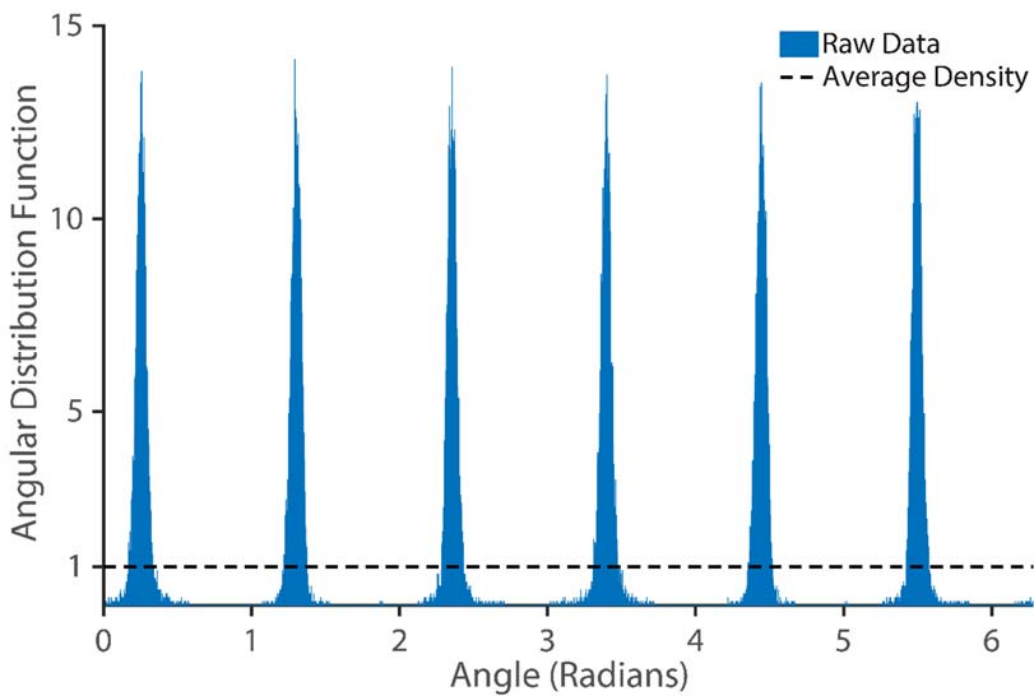

**FIG. S2.** Angular distribution function of the nearest neighbors in the distortion corrected STM image of graphite shown in image of Fig 4(b).

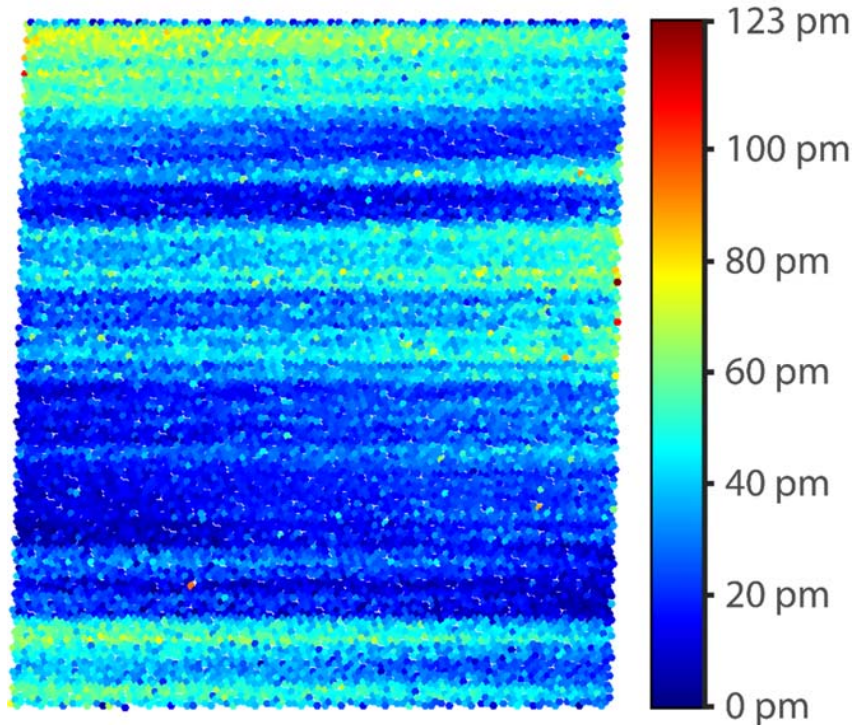

**FIG. S3.** Heat map of the deviation of the features from their whole-image best-fit lattice sites to the 25 nm × 25 nm distortion corrected STM image of graphite shown in Fig. 4(b).

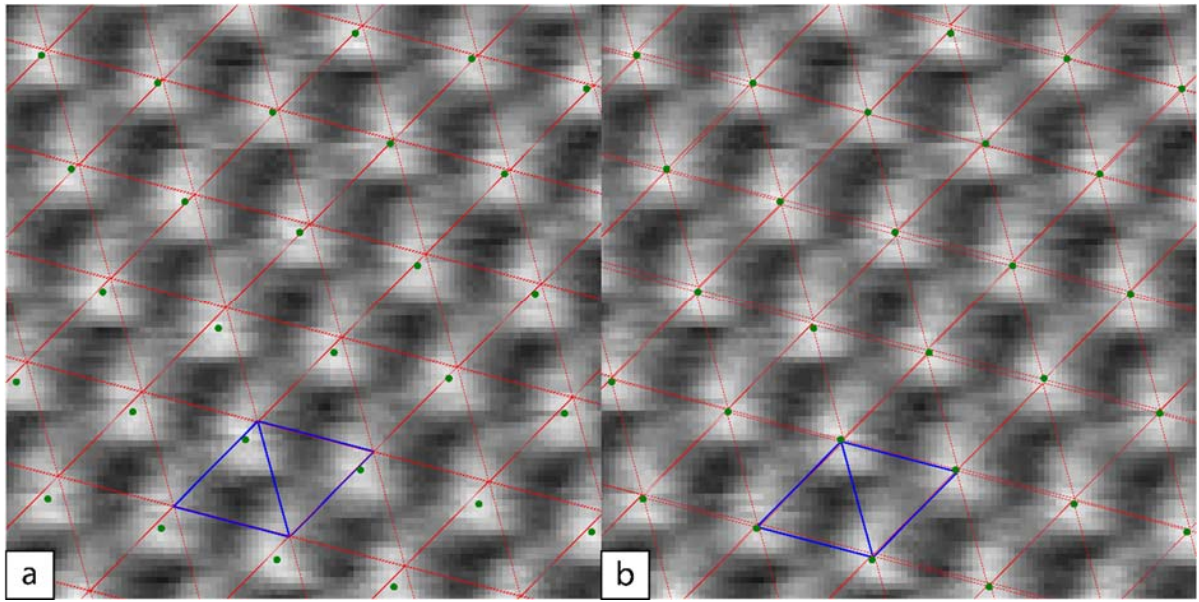

**FIG. S4.** Two identical regions of the distortion corrected graphite image from Fig. 4(b) and the feature locations (green circles) are shown with the whole-image best-fit lattice (a) and the local unit cell fit (b). The fit lattice and units cells are shown in red. An individual unit cell is outlined in blue for clarity.